\documentclass[12pt,preprint]{aastex}

\usepackage{pdflscape}

\slugcomment{To appear in ApJ}

\shorttitle{DY Cen is a binary}
\shortauthors{Rao et al.}

\begin{document}


\title{The hot R Coronae Borealis star DY Centauri is a binary}


\author{N. Kameswara Rao\altaffilmark{1,2}, David L. Lambert\altaffilmark{2},
D. A. Garc\'{\i}a-Hern\'andez\altaffilmark{3,4}, C. Simon Jeffery\altaffilmark{5},
Vincent M. Woolf\altaffilmark{6} and Barbara McArthur\altaffilmark{2}}

\altaffiltext{1}{543, 17$^{th}$ Main, IV Sector, HSR Layout, Bangalore 560102, India; nkrao@iiap.res.in}
\altaffiltext{2}{The W.J. McDonald Observatory, University of Texas, Austin, TX 78712-1083, USA; dll@astro.as.utexas.edu}
\altaffiltext{3}{Instituto de Astrof\'{\i}sica de Canarias, C/ Via L\'actea s/n, 38200 La Laguna, Spain; agarcia@iac.es}
\altaffiltext{4}{Departamento de Astrof\'{\i}sica, Universidad de La Laguna (ULL), E-38205 La Laguna, Spain}
\altaffiltext{5}{Armagh Observatory, College Hill, Armagh, N. Ireland  BT61\,9DG}
\altaffiltext{6}{Physics Department, University of Nebraska at Omaha, NE 68182-0266, USA}


\begin{abstract}
The remarkable hot R Coronae Borealis star DY Cen is revealed to be the first
and only binary system to be found among the R Coronae Borealis stars and their
likely relatives, including the Extreme Helium stars and the hydrogen-deficient
carbon stars. Radial velocity determinations from 1982-2010 have shown DY Cen is
a single-lined spectroscopic binary in an eccentric orbit with a period of 39.67
days. It is also one of the hottest and most H-rich member of the class of RCB
stars. The system may have evolved from a common-envelope to its current form.
\end{abstract}





\section{Introduction}

 DY Centauri may be a remarkable member of the remarkable class of  R Coronae
Borealis (RCB) stars. RCB stars are a rare class of peculiar variable stars with
two principal and defining characteristics: (i) RCBs exhibit a propensity to
fade at unpredictable times by several magnitudes as a result of obscuration by
clouds of dust, and (ii) RCBs have a supergiant-like atmosphere that is very
H-deficient, He-rich and C-rich. The subject of this paper, the remarkable star
DY Cen, is unusual on several accounts, even as a peculiar RCB. DY Cen is one of
the hottest RCBs and it is also one of the most hydrogen-rich RCBs.  DY Cen is
one of only two RCBs to  show features of C$_{60}$ (or proto-fullerenes) in
their spectrum, both being moderately deficient in hydrogen unlike the rest of
the class (Garc\'{\i}a-Hern\'{a}ndez et al. 2011, 2012).  In terms of its
chemical composition, DY Cen may have a composition that sets it apart from most
RCB and Extreme Helium (EHe) stars (see Jeffery \& Heber's (1993) abundance
analysis). DY Cen is not only uncharacteristically H-rich but is  Fe-poor with a
very high S/Fe ratio. However, Jeffery et al. (2011) suggest that the Fe
abundance was greatly underestimated in 1993.

The origins of the hydrogen-deficient stars are still a mystery. Two proposals 
are commonly advocated. In the first proposal, a final helium shell flash occurs on a cooling
white dwarf star  or a very late thermal pulse is experienced by a post-AGB
star  (Iben et al. 1983; Herwig 2000). These events swell the envelope to
supergiant dimensions for a few thousand years and  the remaining hydrogen is
convected inward and consumed while  helium and carbon are convected outward
 to the
surface. Classical examples are believed to be Sakurai's object (V4334 Sgr) and FG
Sge (Asplund et al. 1997; Jeffery \& Sch\"onberner 2006).  The second proposal
involves the merger of two white dwarfs: a carbon-oxygen white dwarf accretes a
helium white dwarf as a close binary orbit shrinks due to energy loss by
gravitational radiation (Webbink 1984; Iben \& Tutukov 1986). The merger leads
to a swollen envelope around the C-O white dwarf which lasts a few thousand
years. Evidence from the chemical compositions of RCB and EHe stars suggests
that most  are products of a merger (e.g., Garc\'{\i}a-Hern\'{a}ndez et al.
2009, 2010; Pandey \& Lambert 2011; Jeffery et al. 2011). The RCB (or EHe) star
from both proposals  will be a single star although a distant companion serving
as a spectator is not ruled out. The possibility of a distant companion in
support of the merger hypothesis seems improbable  on the grounds that triple
stars are rare. 

A search for radial velocity variations of RCB or EHe stars attributable to
orbital motion must face the complication that these stars may exhibit
atmospheric pulsations.  The first systematic radial-velocity surveys of hot
hydrogen-deficient stars (cf. Jeffery et al. 1987)  had revealed evidence of 
 no binaries 
amongst the EHe stars. DY\,Cen was not amongst the sample studied.  However
 individual stars showed pulsation-related
radial motion amongst both RCB stars  (Lawson \& Cottrell
1997) and EHe stars (Jeffery \& Heber 1992; Jeffery et
al. 2001).   DY Cen also showed a variable radial 
velocity     and was suggested to be a possible binary (Giridhar et al. 1996).
 De Marco et al. (2002)  remarked with a striking emphasis that `If
the fact [... that DY Cen could be understood as a spectroscopic binary...] 
could be corroborated,  {\it it would be the first confirmation of binarity in a
RCB star.}' The current  paper which  presents  radial-velocity  measurements
 spread over two decades suggests that DY Cen may in fact be a
spectroscopic binary. 

\section{Spectroscopic Observations}

The general properties of DY Cen's spectrum have been described recently  by De
Marco et al. (2002).  Optical  spectra are  presently a combination  of
photospheric absorption lines (e.g. O\,{\sc ii}, N\,{\sc ii},  Si\,{\sc iii}
lines), emission lines due to a stellar wind  (mainly C\,{\sc ii}, He\,{\sc i}
often superposed on underlying absorption lines,  and nebular emission lines  of
[S\,{\sc ii}], [N\,{\sc ii}], [O\,{\sc i}], [Fe\,{\sc ii}] etc. In addition,
circumstellar and interstellar lines  from Ca\,{\sc ii}, Na\,{\sc i}, K\,{\sc i}
and other species as well as the enigmatic diffuse interstellar bands (DIBs) are
present in absorption (Garc\'{\i}a-Hern\'{a}ndez et al. 2012).

Our first spectra of DY Cen were obtained in 1989 and 1992 with the Cassegrain
 \'Echelle spectrometer at CTIO's Blanco 4m reflector. The resolving power is
close to 18000 for the 1989 spectra and  about 35000 for the 1992 spectrum(
Rao, Giridhar \& Lambert
1993, Giridhar, Rao \& Lambert 1996).

Observations over four years were obtained  at the 3.9m Anglo-Australian
Telescope (AAT) with either the University College London \'Echelle spectrometer
(UCLES)  or the Royal Greenwich Observatory (RGO) spectrograph. The resolving
power of the UCLES spectra is about 40000  and 13000 for the RGO spectra. The
spectral bandpass of the UCLES spectra ran from 4350 \AA\ to 7335 \AA\ in 1999,
3800 \AA\ to 5100 \AA\ in 2002 and 4780 \AA\ to 8800 \AA\ in 2003. The RGO
spectra spanned the interval  3900 \AA\ to  4790 \AA.  These spectra were
retrieved from the observatory's archive and  rereduced.

Spectra of DY Cen were obtained on four nights in February-March 2010 with
the  cross-dispersed \'Echelle spectrograph UVES with the VLT  at ESO's Paranal
observatory. The spectral resolving power as estimated from telluric lines at
6970 \AA\ is 37000.  These spectra cover the regions  3300 to 4500\AA, 5700 to
7525 \AA\ and 7660 to 9460 \AA. The stellar radial velocity was measured from
those  absorption lines longward of  4200 \AA\ which appeared least affected by
emission and asymmetries. The majority of the lines are from O\,{\sc ii}
multiplets and higher ionization species including C\,{\sc iii}, C\,{\sc iv} and
Al\,{\sc iii}. The  centers of the lines have been measured using a cursor. Our
measurements are summarized in Table 1 which also includes   radial velocities 
by Pollacco \& Hill (1991) and  George Herbig who obtained spectra in 1982 using
CTIO's 4-meter Blanco telescope  and a Cassegrain spectrograph. It is expected 
that the  emission line strengths relative to the continuum in 1982 would have
been much weaker. 

In combining velocity measurements from different telescopes and spectrographs,
it is important to check for systematic effects.  A  check in the case of DY Cen
is offered by  the nebular forbidden lines. These originate
from a low density region (Giridhar et al. 1996) which is presumably  remote
from the star and unaffected by its atmospheric changes. Measurements of these
lines  are shown in Table 1, which illustrates that any systematic effects that
exist are within $\pm$ 1.5 km s$^{-1}$. Moreover, the mean of these eleven
measurements is 21.6 km s$^{-1}$, a value consistent with the $\gamma$-velocity
of 21.3 km s$^{-1}$ from the fitted orbit  (Table 2). 

In contrast, the stellar emission lines do vary in velocity. The variation of
emission line velocity seems to be uncorrelated with absorption line velocities.
Moreover the emission is mostly confined to lower multiplets of C\,{\sc ii} and 
He\,{\sc i} lines whereas the measured absorption lines are from higher
ionization and excitation species.

DY Cen was identified as a variable of the RCB type by Hoffleit (1930) from 
well-determined minima in 1897, 1901, 1924 and 1929. No RCB-type minima were
recorded since 1958 (Bateson 1978; A. A. Henden, 2010 private communication). 
In this respect, DY Cen is no longer an active RCB star and all the spectra
discussed here were taken when the star was at maximum light. 

\section{Is DY Cen a spectroscopic binary?}

     For a clear demonstration that DY Cen is a spectroscopic binary, one  should show that the velocity changes do not  arise from atmospheric
disturbances  (e.g., radial and nonradial pulsations or spots). 
Inspection of the cache of spectra shows that many emission lines and also those
absorption lines with emission components  are variable.   While this variability may add
`noise' to the orbital solution, it could provide supporting evidence for the
binary  hypothesis if the spectroscopic variations may be shown to be 
correlated with orbital phase.

In the following subsections, we present the orbital solution, discuss whether
atmospheric pulsations have been mistaken for orbital velocity changes, and
offer comments on line profile variations tied to orbital phase.

\subsection{The spectroscopic binary orbit}

Radial-velocity determinations in Table 1 were examined for periodic variations.
After investigating trial periods between 28 and 50 days, we arrived at a period
of about  39.7 days that gave a satisfactory fit to all measurements. The final
period and the orbital solution were determined using a GaussFit binary model 
(as in  McArthur et al. 2010). Orbital parameters are summarized in Table 2. The
fit to the data is shown in Figure 1 along with  the residuals to that
fit.  The least squares $\chi^2$ of the orbital model was 23.5 for 11 degrees
of freedom (DOF), this indicates that either the errors of the data are
generally underestimated, there are significant outliers in the data set, or the
model is not complete. We used GaussFit's capacity for robust estimation, which
improves the resistance of the parameter values to the effects of `outliers',
which are observations that are not known {\it a priori} to be bad. This is done within the code
by allowing the observations with small residuals to dominate the result,
iteratively reweighting the outliers with higher errors. The $\chi^2$ of the
robust estimation solution is 9.2 for 11 DOF, which indicated that the data set
did contain outliers, which can be seen in the lower portion of Figure 1. In
order to consider the adequacy of the model to fit the observational data, we
also used a sinusoidal model and a linear model to fit the data. The least
squares solution of the sinusoidal model produced a $\chi^2$ of 114.2 for 12 DOF
and the $\chi^2$ of a linear model fit of the data is 144.6 for 14 DOF. These
$\chi^2$ of these two alternative model fits of the data are much higher than
the $\chi^2$ of 23.5 found for the binary orbit model. The sinusoidal model did
not approximate the orbital model because of the significant eccentricity of the
orbital fit. With low eccentricities, the sinusoidal model can approximate the
fit of an orbital model.

The small mass function shows that the unseen secondary has a low mass. The
minimum mass $M_2\sin i$ is around 0.2/$\sin i$$M_\odot$ for, say, a primary
mass of about 0.8$M_\odot$; the secondary mass is insensitive to the assumed
mass of the primary. If $\sin^3 i$ has  its average value (3$\pi/16$ for a
random distribution of inclinations), the secondary mass increases to about
0.3-0.4$M_\odot$. The semi-major axis $a$  is 9.3/$\sin i$ $R_\odot$.   The
secondary may be a low mass helium white dwarf or a low mass (stripped?) main
sequence star. 

\subsection{Atmospheric pulsations?}

Changes in absorption line profiles arising from pulsations or surface spots
translate to a radial velocity change when velocities are measured by our
technique. Fortunately, pulsations, in general, and spots, on all occasions,
will result in a change of line profile. Among the available spectra, only the
VLT/UVES spectra are of the necessary quality to search for subtle line profile
variations. 

Our VLT/UVES spectra sample a phase interval of 0.6. In Figure 2, we
show the mean spectrum spanning the C\,{\sc iv} lines near 5800 \AA\ and
including the DIB at 5796 \AA\ and difference spectra relative to the mean
spectrum. The four  VLT/UVES spectra span  the phase interval between
quadratures (radial velocity maximum and minimum). Individual spectra were
shifted to the C\,{\sc iv} laboratory wavelengths in order to construct the mean
profile and then the difference with the mean spectrum was computed. As Figure 2
shows, these difference profiles around the C\,{\sc iv} lines are featureless
indicating that the absorption profiles over the observing interval were
invariant.  (There is a signal in the difference spectra around the DIB because
this interstellar feature is not subject to the stellar velocity variations.) 

Atmospheric oscillations introduce changes in the observed line symmetry  
 as a velocity shift and line strength  (due to changes in effective
temperature and surface gravity). These changes should be easily visible in the
differenced spectra, especially for temperature sensitive lines such as C\,{\sc iv}.
  None can be seen in the VLT/UVES spectra of DY\,Cen.  
The absence of evidence for oscillations at the few km s$^{-1}$ amplitude in
2010 came as a  surprise because the EHe star BD
$-9^\circ4395$, with atmospheric parameters very similar to those of DY Cen, is
plagued with such oscillations.  BD $-9^\circ$4395 has atmospheric parameters
($T_{\rm eff}$ K,$\log g$ cgs) = (24300,2.65) (Pandey \& Lambert 2011). DY Cen
was measured to have parameters (19500,2.15) by Jeffery \& Heber (1993) but has
evolved to higher temperatures in recent years, say (24800,2.5) (unpublished
analysis).  This is an  almost perfect replica of BD $-9^\circ$4395,  apart
from having approximately 50 times more hydrogen.  Pulsation periods of EHe
stars have been calculated by Saio (2007). Predicted periods for  BD
$-9^\circ$4395 are typically between 2 and 3 days, a range which includes its measured
period of 3.0$\pm1.5$ days for BD $-9^\circ$4395 (Jeffery \& Heber 1992). The
scale of the line profile variations for BD $-9^\circ4395$ corresponds to a
dispersion profile of peak-to-peak amplitude of about {\bf $\pm0.05$ } of the
continuum flux in  differenced spectra of strong lines like the C\,{\sc iv}
lines in Figure 2 (see Fig. 7 in Jeffery \& Heber 1992).    Such features are
clearly not present in difference spectra for DY Cen.
 
The contrast between the presence of nonradial oscillations in BD $-9^\circ4395$ and their absence in DY Cen in 2010 is surprising. Perhaps, their
excitation is intermittent or even quenched at the higher H abundance of DY Cen:
the H abundance is $\log \epsilon$(H) = 10.8 for DY Cen (Jeffery \& Heber 1993)
and 9.1 for BD $-9^\circ4395$ (Pandey \& Lambert 2011).
Hydrogen quenching is evident in pulsation models for higher gravity EHe stars, only
when the hydrogen mass fraction exceeds 30\%  (Jeffery \& Saio 1999); consequently
DY Cen should still be unstable to radial pulsations.

                                            Pollacco \& Hill
(1991) reported low-amplitude ($\Delta$V$\sim$ 0.1) short-period (3.8 - 5.5
days) photometric variations from 1987 observations.  Surface pulsations
producing such light variations  would be associated  with
velocity variations  of up to  20 km\,s$^{-1}$ (see 
Jeffery et al. 2001). Since we cannot exclude the presence of
such motion from, at least, the earlier observations in our dataset, we have 
to acknowledge
that our current orbital solution is uncertain to at least this extent.

        The 1987 observations also showed a  longer duration ($<25$ days) and
larger amplitude  ($\Delta$ V $\sim$ 0.2) and appeared to confirm reports from
Bateson (1978) from visual observations of a longer period  variation.
Such a variation may not be intrinsic to the star but may  arise from changing
extinction as the star moves in its orbit. In this regard, one recalls the
behaviour of RV Tauri variables and the photometric distinction  between the RVa
and RVb types. 

\subsection{Phase-dependent spectroscopic changes?}

Our  optical spectra of DY Cen show that there are short and
long term changes in the spectral features. 
There is a tantalising hint of orbital phase dependent changes in the 4267 \AA\
C\,{\sc ii}  emission profile - see Figure 3. Near conjunction, 
 one or more absorption components appear superimposed on the emission
profile. (A similar profile is also seen at the other conjunction (primary on the
farside) from the observations obtained in 2002 June 26.)  However, the profile
of the C\,{\sc ii} 7231 \AA\ emission line is essentially  the same on all four
spectra with no hint of superposed absorption,   and the C\,{\sc ii} 6578 \AA\
line shows a P Cygni profile at all four phases indicative of strong mass loss.
Additional observations with complete phase coverage will be needed to
investigate phase-dependent spectral variations.

\section{Concluding Remarks}

The variable radial velocities of DY Cen suggest that it is a single-lined
spectroscopic binary. The  absorption line profiles over the interval spanned by
the VLT/UVES observations indicate that nonradial oscillations are  unlikely
to account for the amplitude of the radial-velocity variation in 2010.  Thus, we
assume that DY Cen is a product of binary star evolution, unlike all other
members of RCB class.

Consideration of the size of the orbit and  the radius of DY Cen shows that the
components of this binary have experienced and are likely to continue to experience mass
loss and mass transfer which is revealed by the presence of emission lines
throughout the spectrum.  For example, the  semi-major axis a$\sin i$ =
9.3$R_\odot$ may be compared with the  radius of DY Cen.
Assuming a mass of 0.8 M$_\odot$ and using the recent measurement
$\log g=2.5$  gives a stellar radius $\approx 8$ R$_\odot$.
Looking back 20 years and using $\log g = 2.15$ (Jeffery \& Heber 1993) gives a 
radius $\approx 12$ R$_\odot$.
Considering the eccentricity $e=0.44$ given by our orbital solution,
periastron passage will occur with a separation of  some $10/\sin i$ R$_\odot$. 
This close proximity and evidence that DY Cen is contracting at a substantial rate
may well explain both the very active emission-line activity and the peculiar
nature of DY\, Cen in the recent past. If the surface of DY Cen had been as cool as 5\,000\,K,
its radius would have exceeded 100 R$_\odot$, several times the present orbital separation.
Indeed, {\it it is possible that DY\,Cen was
a true common-envelope system, with the secondary embedded within the envelope
of the primary, within the last 100 years.}
Even this episode was almost certainly preceded by earlier
mass transfer and orbital changes; the radius of a normal star  on the
first giant branch is  greater than the inferred semi-major
axis -- unless the angle of inclination is very high.

As a close binary with a  substantial surface abundance of hydrogen, and with an apparent contraction
rate  in excess of any observed amongst the EHe and classical RCB stars, the latter are probably not useful types
for considering the evolution of DY\,Cen. Similarly, the more massive hydrogen-deficient
binaries  including $\upsilon$\,Sgr and KS\,Per are  extremely hydrogen-deficient and nitrogen-rich and unlikely counterparts.
           If DY\,Cen continues to contract at the rate
suggested by the  measurements of $T_{\rm eff}$ and $\log g$ (Jeffery \& Heber 1993 and
 the present),
it is possible that there is no known  star in the Galaxy quite like it.

Regarding its future evolution one can do little more than speculate.
Although at present DY Cen is probably transferring mass to its companion,
its rapid contraction implies that mass transfer will soon cease.
That DY\,Cen will become first a hot subdwarf and subsequently a white dwarf appears to be clear.
Less clear is whether it presently has a  helium or a  carbon-oxgen core. The first case should lead to
a stable helium-core burning phase, where it might be identified amongst the intermediate helium
subdwarf stars (cf. CPD$-20^{\circ}1123$, Naslim et al. 2012).
Since atomic diffusion transforms the surface composition of subdwarf B stars, it is difficult to identify possible successors reliably;
intermediate helium subdwarfs come with a range of surface carbon and nitrogen abundances
for example ($0.1 < {\rm C/N} < 10$, Naslim et al. in prep).
It is not known whether intermediate helium stars become classical sdB stars -- which are very {\it helium deficient}.
Since the processes that form sdB stars in binaries  remove most of the hydrogen envelope, one might expect
their surfaces to be somewhat hydrogen-deficient until radiatively-driven diffusion forces the helium to sink below sight,
so DY\,Cen could be an immediate post-common-envelope binary in transition to the subdwarf B phase. The majority of sdB stars
are binaries, and several have periods in the 10 -- 100 day range (Maxted et al. 2001; Copperwheat et al. 2011; Barlow et al. 2012).
On the other hand, if it is slightly more massive, and already has a carbon-oxygen core, then it will evolve
directly through the helium-rich subdwarf O star and on to become a hot white dwarf. Its subsequent appearance
and evolution will depend on the nature of the unseen secondary.



\acknowledgments

We acknowledge the anonymous referee for suggestions that help to improve the
paper. This work is based on observations obtained with the ESO programme 284.D-
5048(A). D.A.G.H. acknowledges support  provided by the Spanish Ministry of
Economy and Competitiveness under grant AYA$-$2011$-$27754. Our sincere thanks
are due to George Herbig for supplying his radial velocity measurements. We 
acknowledge with thanks use of SIMBAD and AAVSO databases.



{\it Facilities:} \facility{VLT: Kueyen}.

\clearpage

\begin{landscape}

\begin{deluxetable}{rrrrrrll}
\tabletypesize{\scriptsize}
\tablecaption{Radial velocity measurements of DY Cen}
\tablewidth{0pt}
\tablehead{
\colhead{Date} & \colhead{Julian Date} & \colhead{Phase} & \colhead{Rad.Vel.} & \colhead{No.} & \colhead{Emission$^a$} & \colhead{Nebular} &\colhead{Source} \\
\colhead{(UT)} & \colhead{(2400000+)} & \colhead{} & \colhead{km s$^{-1}$} & \colhead{lines} & \colhead{Lines(km s$^{-1}$)} & \colhead{Lines(km
s$^{-1}$)} &  \colhead{}
}
\startdata
1982 Apr 9  &45068.734&0.102  &35  $\pm$3   &    &    &         & Herbig\\
1982 Apr 11 &45070.624&0.150  &25  $\pm$3   &    &    &         & Herbig\\
1982 Apr 11 &45070.664&0.151  &25  $\pm$3   &    &    &         & Herbig\\
1982 Apr 12 &45071.694&0.177  &32  $\pm$3   &    &    &         & Herbig\\
1988 Mar 13 &47233.774&0.682  &15.1$\pm$2.5 &    &    &         & Pollacco \& Hill\\
1989 July 16 &47723.70&0.032  &41  $\pm$4    &  &-2.4$\pm$4.0 &22.8$\pm$1.9 & Rao et al.\\
1992 May 20 &48762.608&0.223  &29  $\pm$4    &  &26$\pm$6&21.5$\pm$2.4 & Giridhar et al.\\
1999 Feb 5  &51215.248&0.052  &37.3$\pm$2.4 & 21 & 23.3$\pm$1.1& 23.7$\pm$1.1& AAT/UCLES\\
2001 Apr 4  &52003.998&0.936  &31.7$\pm$2.7 &    & 15.0$\pm$3.2&21.7$\pm$2.1& AAT/RGO\\
2001 Apr 5  &52005.137&0.964  &32.5$\pm$2.6 &    & 17.4$\pm$2.5&19.6$\pm$2.3& AAT/RGO\\
2002 Jun 26 &52449.926&0.178  &23.8$\pm$2.0 & 19 & 20.1$\pm$1.5&20.1$\pm$ & AAT/UCLES\\
2003 Jul 18 &52838.903&0.984  &40.0$\pm$1.5 & 11 & 14.0$\pm$5.0&22.3$\pm$1.2& AAT/UCLES\\
2010 Feb 27 &55254.772&0.886  &24.7$\pm$2.6 & 36 & 14.8$\pm$2.6&21.3$\pm$1.3& VLT/UVES\\
2010 Mar 2  &55257.749&0.961  &25.4$\pm$2.5 & 47 & 15.7$\pm$3.1&20.9$\pm$1.1& VLT/UVES\\
2010 Mar 5  &55260.795&0.038  &36.9$\pm$4.6 & 34 & 24.7$\pm$2.7&20.9$\pm$1.3& VLT/UVES\\
2010 Mar 25 &55280.601&0.537  &14.2$\pm$3.6 & 28 & 22.8$\pm$3.3&20.6$\pm$1.1& VLT/UVES\\
\enddata
\end{deluxetable}
\end{landscape}

\clearpage

\begin{deluxetable}{ll}
\tabletypesize{\scriptsize}
\tablecaption{Orbital parameters of DY Cen}
\tablewidth{0pt}
\tablehead{
\colhead{Parameter} & \colhead{Value} 
}
\startdata
 T$_{\rm o}$ (JD) & 2445104.3364$\pm$ 1.715\\
 P &     39.66779$\pm$0.0088 d \\
 $\gamma$ &    21.30$\pm$0.45 km s$^{-1}$\\
 K &   13.26$\pm$1.18 km s$^{-1}$ \\
 e  &  0.44$\pm$0.10\\
 $\omega$ &   344.5$\pm$16.7\\
 f(M) &6.92$^{+0.43}_{-0.91}$ 10$^{-3}$ M$_\odot$\\
\enddata
\end{deluxetable}

\clearpage

\begin{figure}
\includegraphics[angle=0,scale=.65]{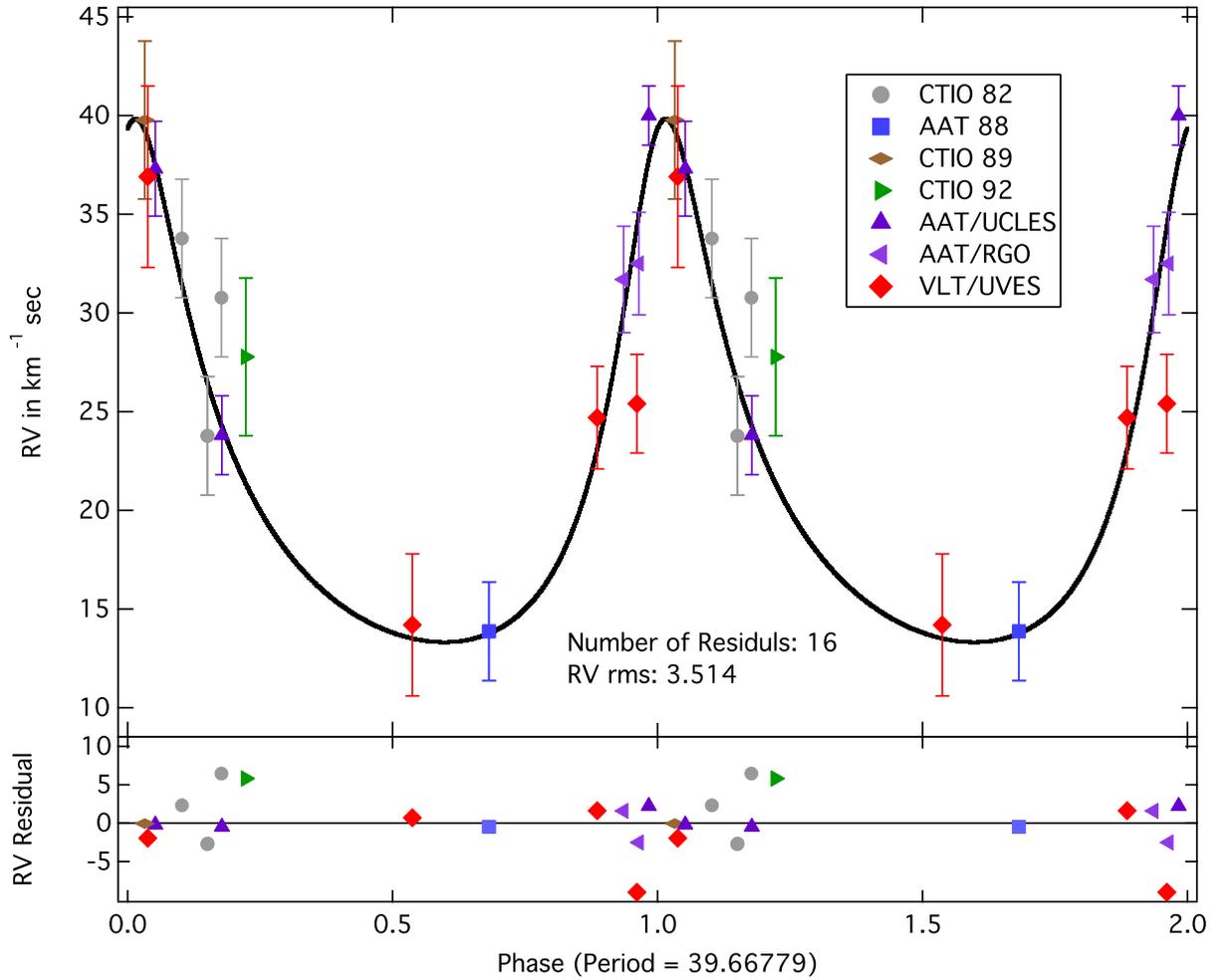}
\caption{Stellar radial velocity curve similar to Figure 3. The deviations of
observations to the computed fit for the parameters in Table 4 are shown in the
bottom.The notation of the observations is shown in the inset. \label{fig1}}
\end{figure}

\clearpage

\begin{figure}
\includegraphics[angle=0,scale=.65]{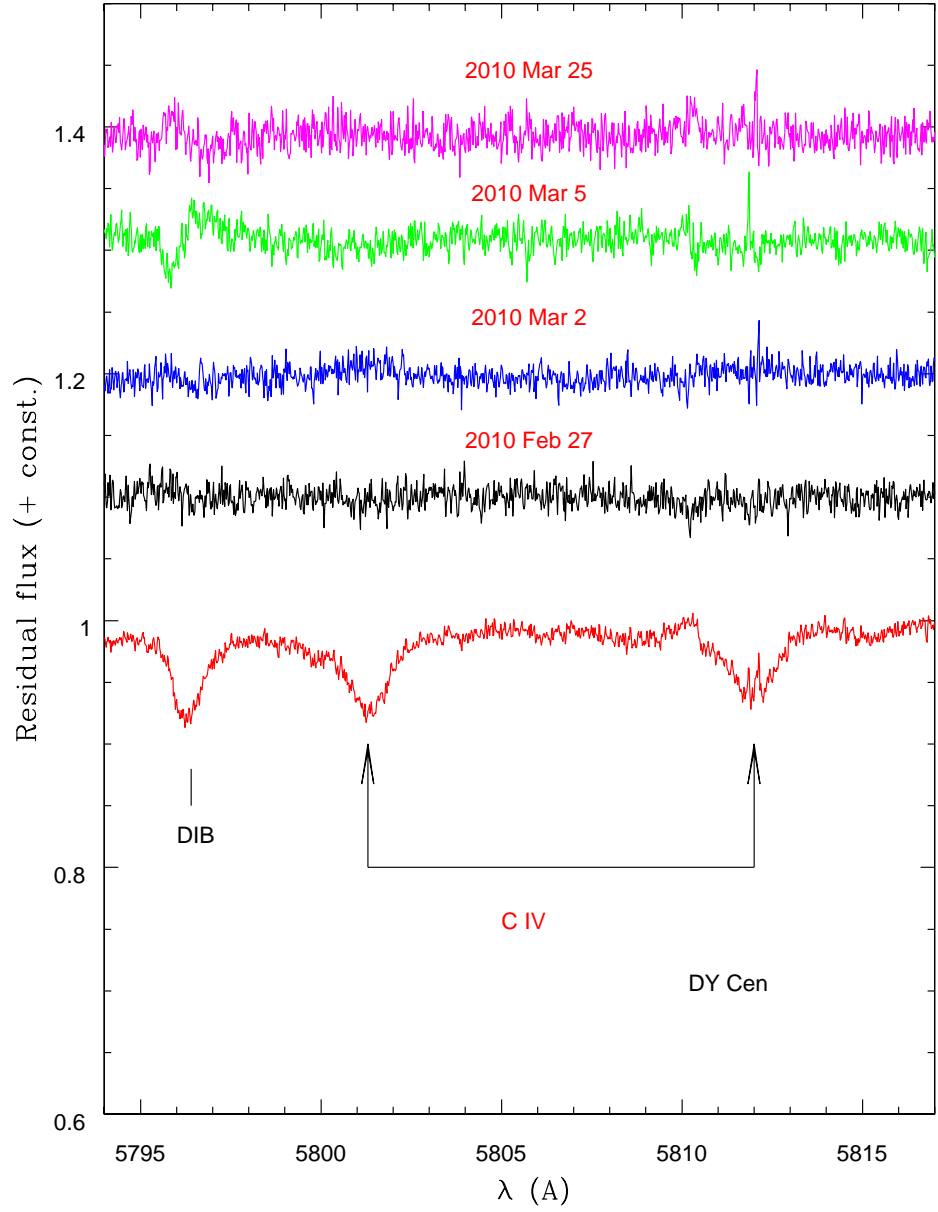}
\caption{Mean and differenced spectra around the C\,{\sc iv} 5800-5810 \AA\
lines. The sense of the differenced spectra is `observed - mean'.\label{fig2}}
\end{figure}

\clearpage

\begin{figure}
\includegraphics[angle=0,scale=.65]{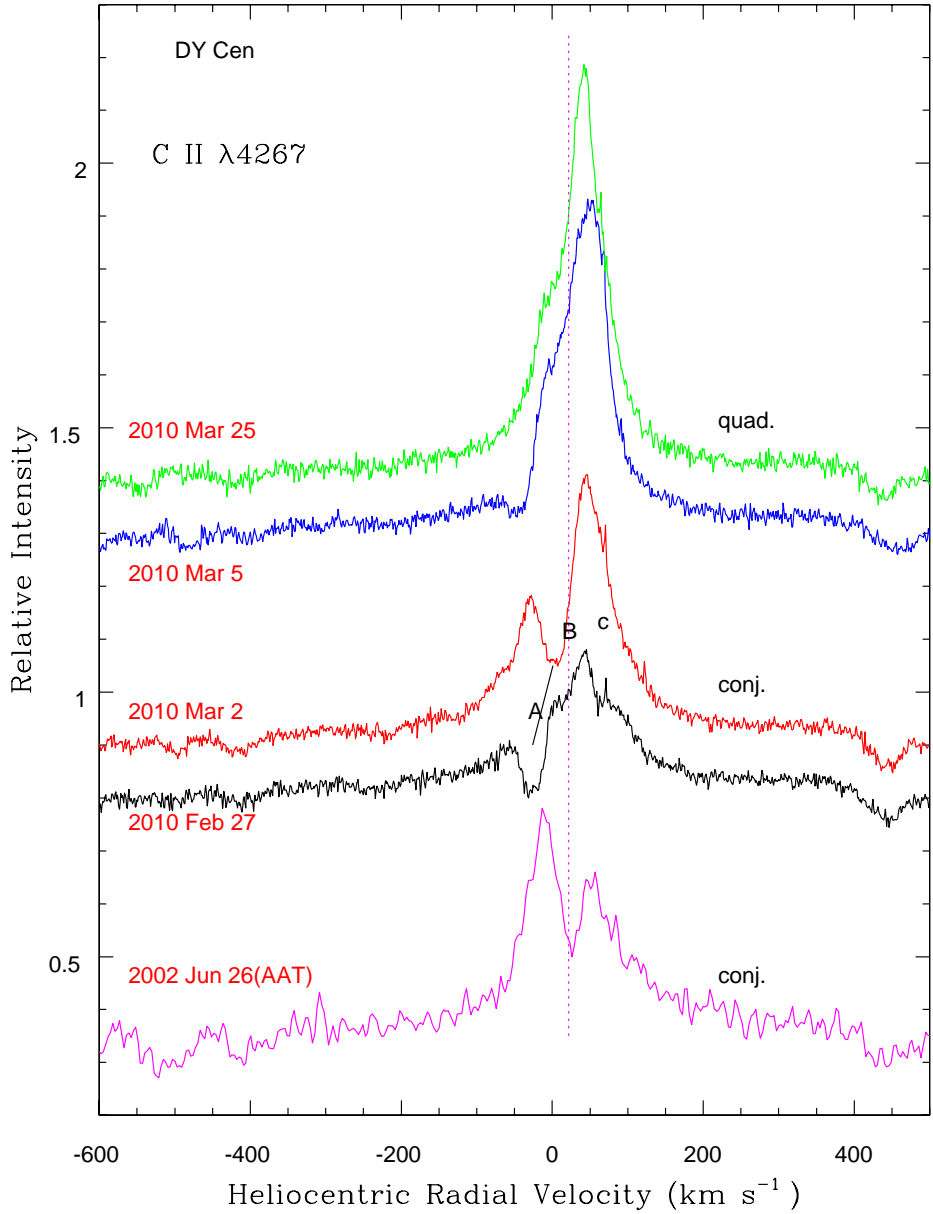}
\caption{C\,{\sc ii} 4267 \AA\ profiles during
the 2010 February--March period. The spectra are shifted by orbitrary  amounts
in y-axis for clarity.  Absorption components  are superposed on the
broad emission profile. These components are prominent in the profiles
 obtained near  conjunction. The
emission flux is lowest near conjunction. Vertical dashed line represents the
systemic velocity (obtained from nebular lines).\label{fig3}}
\end{figure}

\end{document}